\documentclass[12pt]{spieman}  
\usepackage{amsmath,amsfonts,amssymb}
\usepackage{graphicx}
\usepackage{setspace}
\usepackage{tocloft}
\usepackage[colorlinks=true, allcolors=blue]{hyperref}
\usepackage{comment}
\usepackage{multirow}
\usepackage{multicol}
\usepackage{subcaption}
\usepackage[normalem]{ulem}
\usepackage{textcomp}
\usepackage{tabularx}  

\title{Prototype sub-wavelength structure anti-reflection coating on alumina filters for ground-based CMB telescopes}

\author[a,*]{Kosuke Aizawa}
\author[a]{Ryosuke Akizawa}
\author[b]{Scott Cray}
\author[b]{Shaul Hanany}
\author[c]{Shotaro Kawano}
\author[d]{J\"{u}rgen Koch}
\author[c]{Kuniaki Konishi}
\author[b]{Rex Lam}
\author[e,f,g,h]{Tomotake Matsumura}
\author[c]{Haruyuki Sakurai}
\author[i]{Ryota Takaku}

\affil[a]{Department of Physics, The University of Tokyo, 7-3-1 Hongo, Bunkyo-ku, Tokyo 113-8654, Japan}
\affil[b]{School of Physics and Astronomy, University of Minnesota Twin Cities, 115 Union St. SE, Minneapolis MN 55455, USA}
\affil[c]{Institute for Photon Science and Technology (IPST), The University of Tokyo, 7-3-1 Hongo, Bunkyo-ku, Tokyo 113-0033, Japan}
\affil[d]{Laser Zentrum Hannover e.V., Hollerithallee 8, 30419 Hannover, Germany}
\affil[e]{Kavli Institute for the Physics and Mathematics of the Universe (IPMU), The University of Tokyo, 5-1-5 Kashiwa-no-Ha, Kashiwa, Chiba 277-8583, Japan}
\affil[f]{Center for Data Driven Discovery (CD3), Kavli Institute for the Physics and Mathematics of the Universe (IPMU), The University of Tokyo, 5-1-5 Kashiwa-no-Ha, Kashiwa, Chiba 277-8583, Japan}
\affil[g]{ILANCE, CNRS, University of Tokyo International Research Laboratory, Kashiwa, Chiba 277-8582, Japan}
\affil[h]{Institute of Space and Astronautical Science (ISAS), Japan Aerospace Exploration Agency (JAXA) 3-1-1 Yoshinodai, Chuo, Sagamihara, Kanagawa 252-5210, Japan}
\affil[i]{Department of Physics, Okayama University, 1-1-1 Tsushimanaka, Kita-ku, Okayama city, Okayama, 700-8530, Japan}

\begin{document}

\renewcommand{\cftdotsep}{\cftnodots}
\cftpagenumbersoff{figure}
\cftpagenumbersoff{table} 
\maketitle

\begin{abstract} 

We present designs and fabrication of sub-wavelength anti-reflection (AR) structures on alumina for 
infrared absorptive filters with passbands near 30, 125, and 250~GHz. These bands are widely used by ground-based  instruments measuring the cosmic microwave background radiation. The designs are tuned to provide reflectance of 2\% or less for fractional bandwidths between 51\% and 72\%, with each of the three primary bands containing two sub-bands. We make the sub-wavelength structures (SWS), which resemble a two-dimensional array of pyramids, using laser ablation. We measure the shapes of the fabricated pyramids and show that for incidence angles up to 20~degrees  the predicted in-band average reflectance is 2\% or less, in agreement with the design. The band average instrumental polarization is less than $3\times 10^{-3}$.


\end{abstract}

{\noindent \footnotesize\textbf{*}Further author information: K. Aizawa: E-mail:  \linkable{kosuke.aizawa@ipmu.jp} }


\section{INTRODUCTION}

The optical chain of cosmic microwave background (CMB) receivers contain filters that reflect or absorb incident radiation. 
High quality absorptive filters have low in-band and high out-of-band absorption, high thermal conductance, and low reflectance. 
Alumina (polycrystaline Al$_{2}$O$_{3}$) is an attractive material~\cite{Inoue2014}. 
At cryogenic temperatures alumina's loss tangent (tan$\delta$) is $\sim 10^{-4}$~\cite{Inoue2014}, among the lowest compared to other absorptive materials, and its thermal conductivity is $\sim1~$W/cm$\cdot$K at 50~K~\cite{NEMOTO1985531}, which is about 10 times higher than teflon's~\cite{Choi2012}, a material that has also been used as a millimeter-wave filter~\cite{2010:BICEP2,2012:POLARBEAR,2015:BICEP2/KECK}.   
However, alumina's high index of refraction $n \simeq 3.1$\cite{Lamb1996} requires the implementation of anti-reflection coating (ARC) without which broad-band reflective losses could exceed 40\%.  

Several techniques have been proposed to make ARC on materials used in millimeter-wave optics~\cite{Raut2011,Hargrave2010,Wheeler2014,Lau:06,Rosen:13,Inoue2016,Nadolski:20,Sakaguri2024,Jeong:23,Datta:13,Nitta2017}.
For ceramics, we have been developing ARC based on SWS patterned  directly on the surface of the material~\cite{matsumura2016,Schutz2016,Young2017,matsumura2018,Sakurai2019,takakuJAP2020,takakuSPIE2020,wen2021,mustang2,Takaku2022,Golec:22,Takaku2023}. Because alumina has hardness 9 on the Mohs scale~\cite{Frost1983} and direct machining is challenging, we are using laser ablation to pattern the SWS~\cite{matsumura2016,Schutz2016,takakuSPIE2020,wen2021,mustang2}.  

Ground-based CMB instruments typically operate in several atmospheric windows between 20 and 300~GHz \cite{bicep/keck,SPTpol,SO2019,Ghosh_2022,PB2-A,barron2022review}.
In this paper we present designs and prototypes for SWS-ARC for alumina filters for such windows.
To use a concrete example, we focus on the bands of the Simons Observatory's small aperture telescope (SO-SAT)~\cite{SO_SAT:2024}. 
In Section~\ref{sec:bands} we review the bands for which the ARC are designed. In section~\ref{sec:design},~\ref{sec:fabrications}, and~\ref{sec:results} we discuss the design of the SWS, the fabrication of prototypes, and we give estimates for full sample reflectance and instrumental polarization, respectively. We discuss and summarize in Section~\ref{sec:discussions}. While the paper focuses on the SO-SAT bands, the methodology we present is general and can be extended to other instruments and frequency bands.

\section{Frequency Bands} \label{sec:bands}

Table~\ref{tab:band} lists the bands of SO-SAT~\cite{SO_SAT:2024}. There are three primary bands, a Low, Middle, and High Frequency bands (LF, MF, and HF), and each has two sub-bands. All sub-bands have a fractional bandwidth of 30\%\cite{Sakaguri2024}.
The light path of each primary band includes one alumina-based filter shared among the two sub-bands, therefore requiring high efficiency ARC over fractional bandwidths up to 72\%, see Table~\ref{tab:band}. We present three SWS topologies, one per band, that are designed to give average reflection less than 0.02 per primary band. 

\begin{table}[ht]
    
        \caption{Primary frequency bands, sub-bands, and the required bandwidths. The fractional bandwidth of each sub-band is 30\%~\cite{Sakaguri2024}. 
        } 
        
        \resizebox{\textwidth}{!}{
        
            \begin{tabular}{ c  c  c  c  c  c  c }
    
            \hline
            Primary band & 
            \multicolumn{2}{c}{LF} & \multicolumn{2}{c}{MF} & \multicolumn{2}{c}{HF} \\
            \hline
            Center freq. [GHz] & 27 & 39 & 93 & 145 & 225 & 280 \\
            \hline
            (Min. freq., Max. freq.) [GHz] & (23, 31) & (33, 45) & (79, 107) & (123, 167) & (191, 259) & (238, 322) \\
            \hline
           Fractional bandwidth & \multicolumn{2}{c}{0.65} & \multicolumn{2}{c}{0.72} & \multicolumn{2}{c}{0.51} \\
            \hline
            \end{tabular}
        
        }
        \label{tab:band}
\end{table}

\section{SWS Design}
\label{sec:design}

The ARC is assumed to be a periodic array of structures with period, or pitch $p$. 
We assume an index of refraction $n_{\rm{sub}}=3.12$~\cite{Lamb1996,mustang2}, and because we focus on reflection properties all calculations include no loss, $\tan{\delta}=0$.  The array of approximately pyramidal-shaped structures of height $d$ covers both sides of a slab of thickness  $t=3.0$~mm. This is the value specified for the MF and HF filters of the SO-SAT~\cite{Sakaguri2024}.


\subsection{Pitch Determination}

The value of the pitch is determined by requiring no diffraction within the passband of the highest frequency band~\cite{Grann:95,Bruckner:07} 
\begin{eqnarray}
        p \leq \frac{c/\nu_{\rm{h}}}{n_{\rm{sub}}+\sin{\theta_{\rm max}}},
\end{eqnarray}
where $c$ is the speed of light, $n_{\rm{sub}}$ is the index of alumina, $\nu_{h}$ is the highest passband frequency per primary band as given in Table~\ref{tab:band},  and $\theta_{\rm max}$ is the largest incidence angle of the incoming radiation, which for SO-SAT is $\theta_{\rm max} = 17.5$~degrees taking as the half angle of the field-of-view~\cite{SOSAT_2020}.
The calculated pitch values for each primary band are given in Table~\ref{tab:designed_pitch_depth}.
For an actual filter to be implemented with an instrument, it would be appropriate to use a frequency $\nu_{h}$ that is higher than the highest pass-band frequency. 


\begin{table}[ht]
    \centering
    \caption{The designed pitch $p$ and depth $d_{\text{opt}}$ of the SWS for each primary band.}
    \begin{tabular}{ c  c  c  c } 
    \hline
     & LF & MF & HF \\
    \hline
    \hline
    $p$ [$\mathrm{mm}$] & 1.90  & 0.50  & 0.27   \\
    \hline
    $d_{\text{opt}}$ [$\mathrm{mm}$] & 4.45  & 1.31  & 0.55   \\
    \hline
    \end{tabular}
    \label{tab:designed_pitch_depth}
\end{table}

\subsection{SWS Shape Design}

Klopfenstein has proposed an optimal impedance matching method to minimize the reflection coefficient in a transmission-line application. The approach relies on tapering the impedance~\cite{Klopfenstein1956}. Given a frequency band, the design is determined by two parameters: the length of the taper $d$, and $\Gamma_{\rm{m}}$, which determines the in-band ripple. We apply his method to find an optimal shape for each of LF, MF, and HF primary frequency bands.~\cite{Young2017, matsumura2018,takakuJAP2020,mustang2} 

\begin{figure}[ht]
    \centering

    \includegraphics[width = 0.4\textwidth,height=0.25\textwidth]{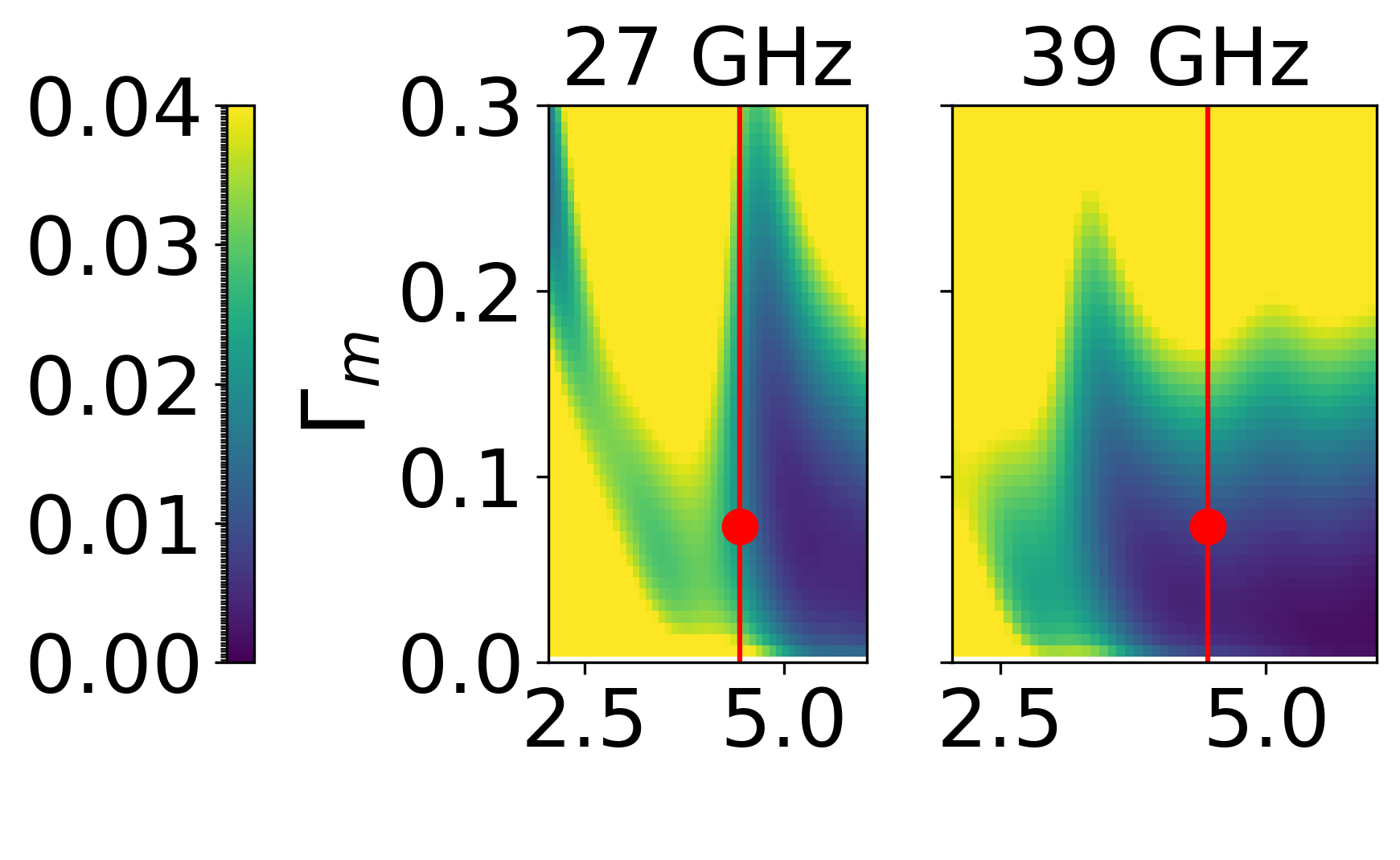}\vline
    \includegraphics[width = 0.25\textwidth,height=0.25\textwidth]{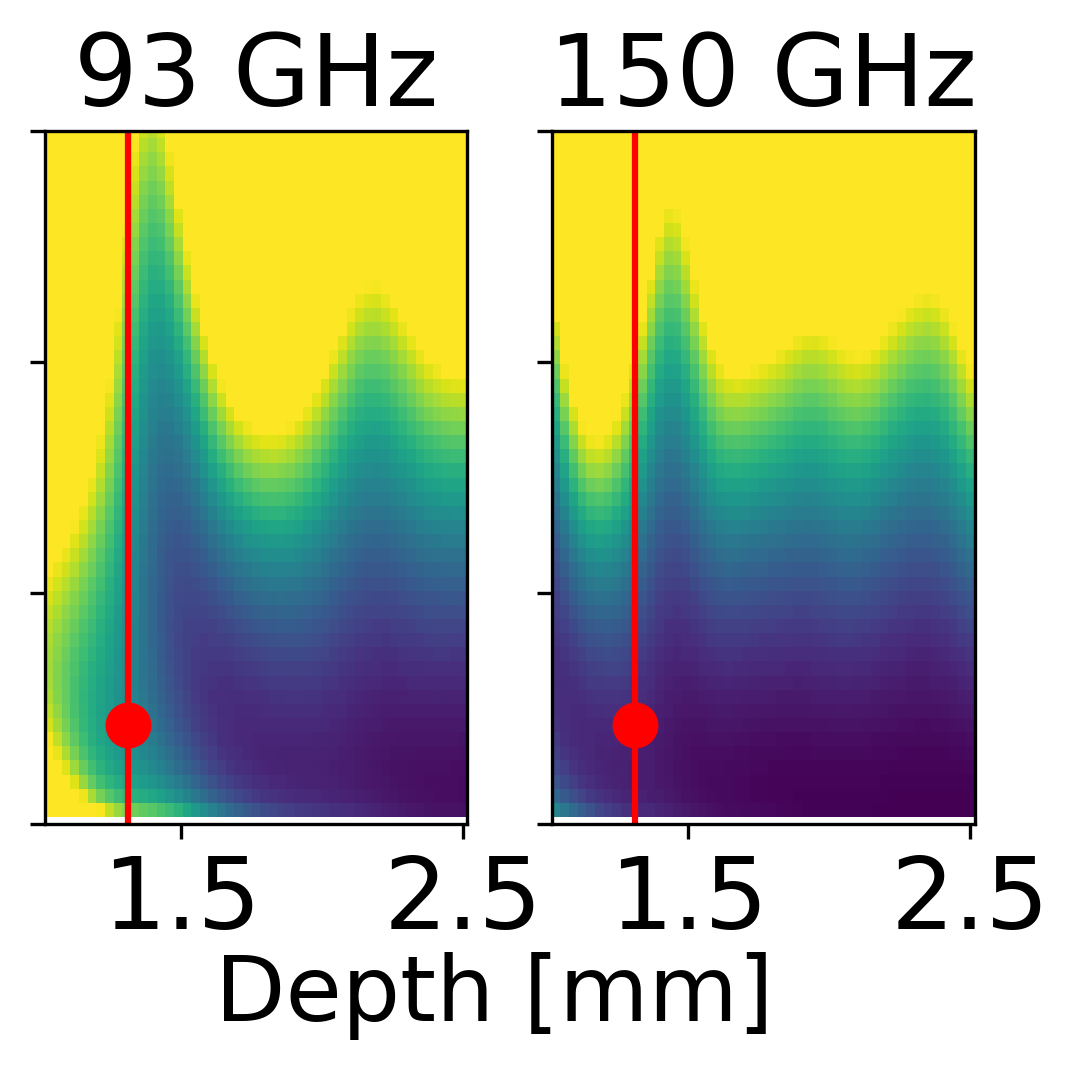}\vline
    \includegraphics[width = 0.25\textwidth,height=0.25\textwidth]{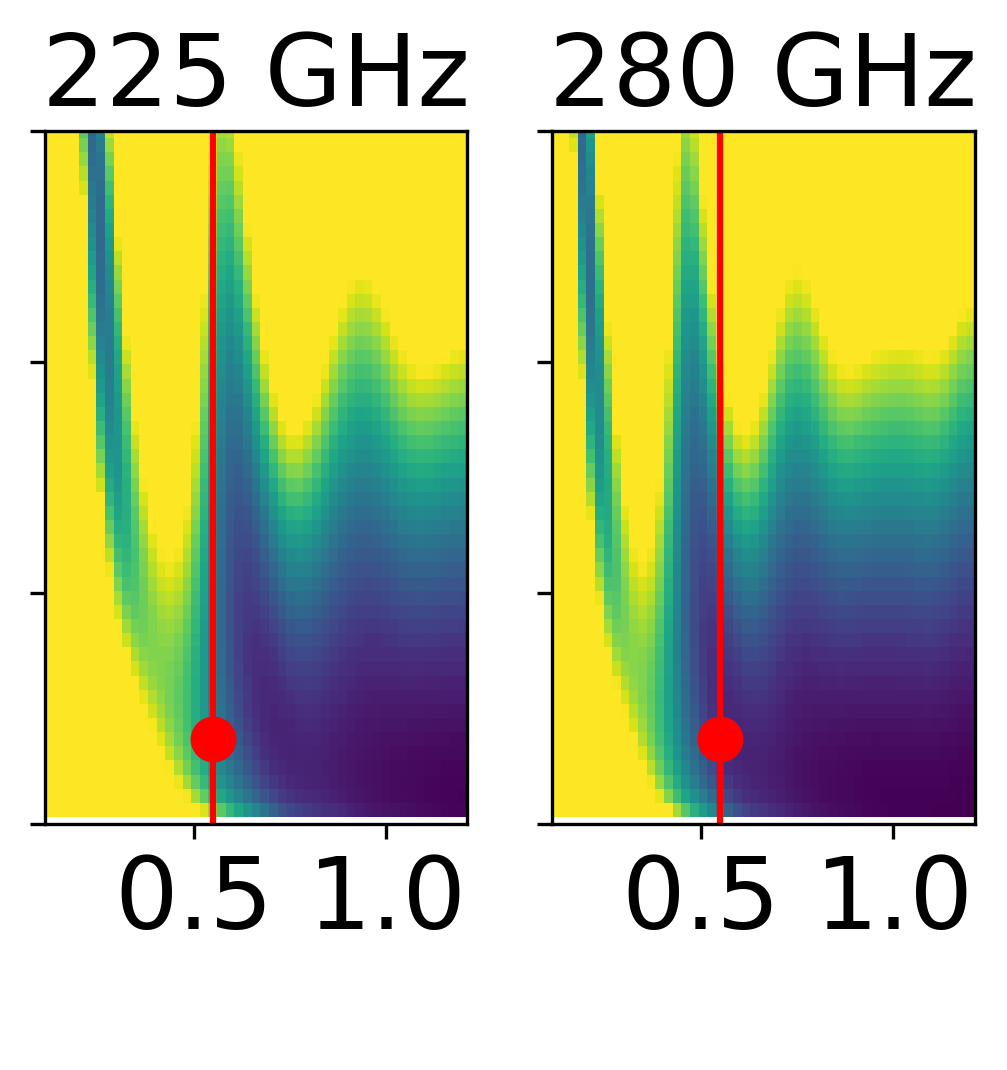}

    \includegraphics[width = 0.48\textwidth]{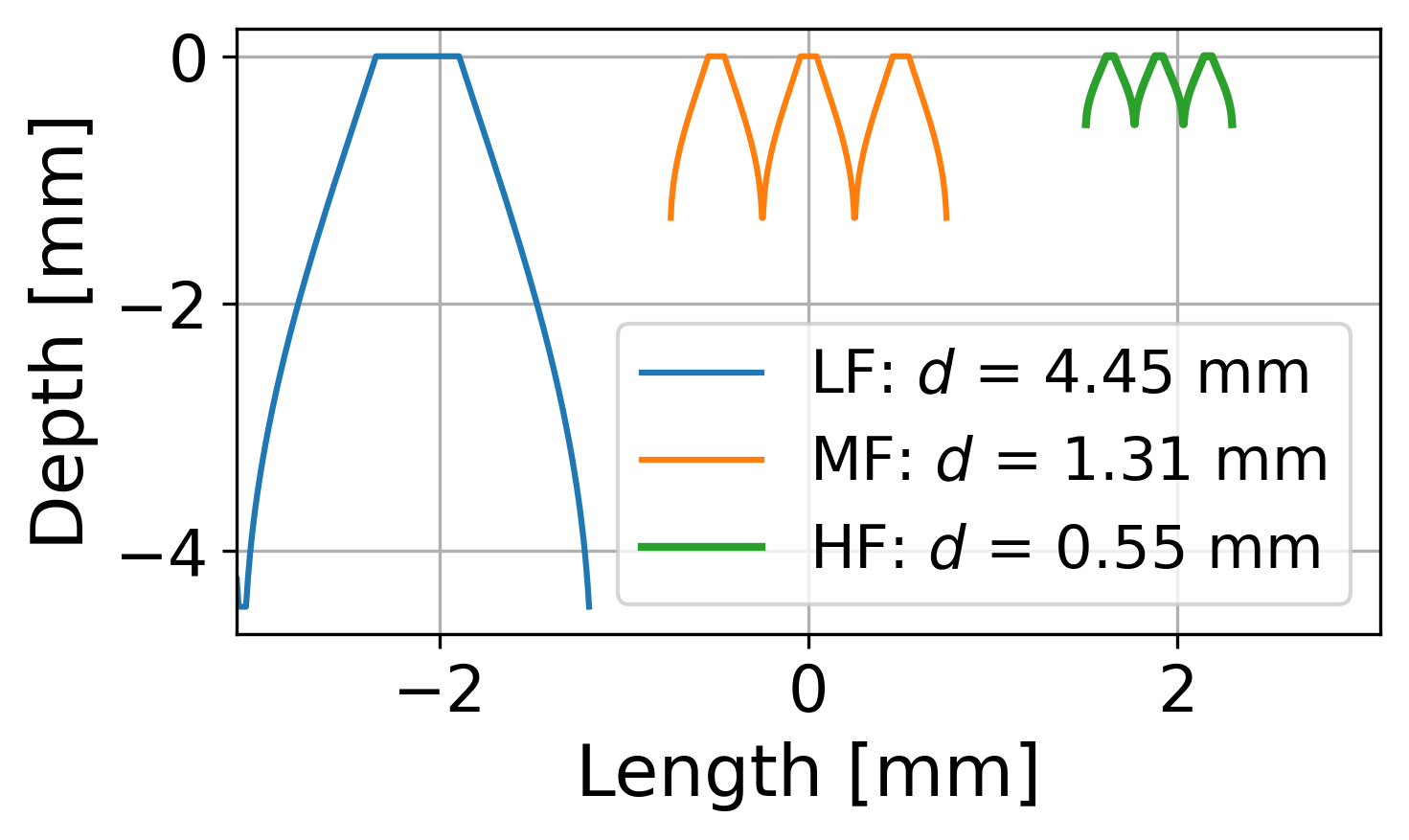}
    \includegraphics[width = 0.48\textwidth]{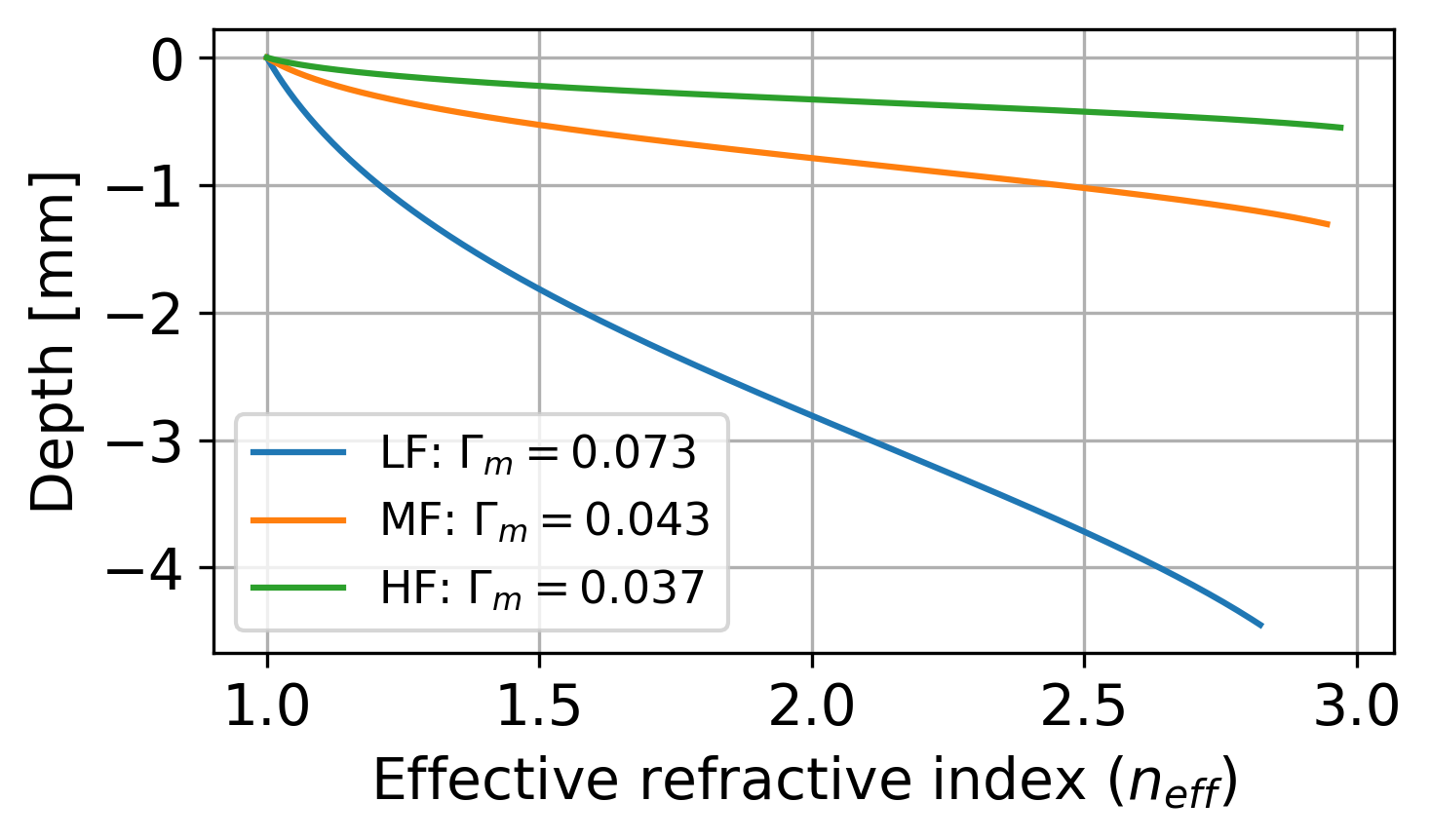} \\

    \includegraphics[width = 0.9\textwidth,height=0.18\textwidth]{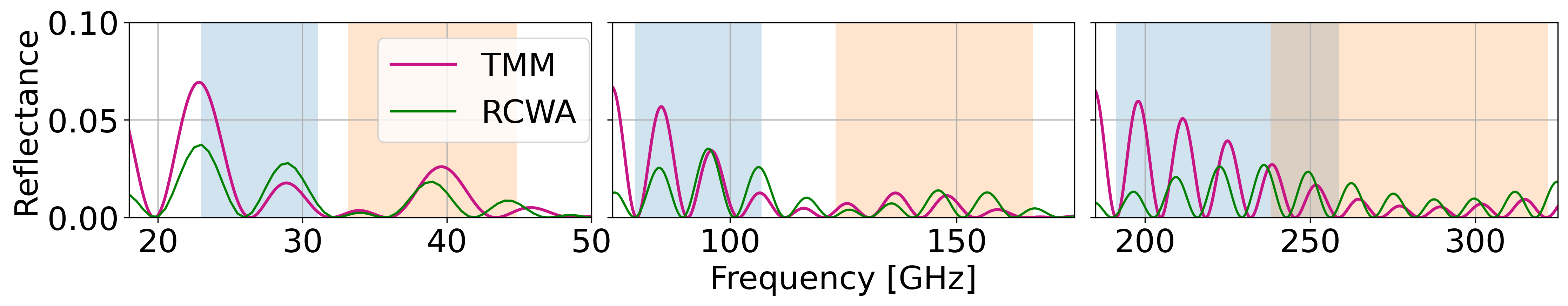}

    \caption{First row: Averaged reflectance $\bar{R}_{l}$ and $\bar{R}_{h}$ for the lower and higher sub-band, respectively (color map), as a function of depth $d$ and $\Gamma_{\rm{m}}$, the values of $d_{\text{opt}}$ (red vertical line) and the chosen $\Gamma_{\rm{m}}$ (red dot). 
    Second row: The selected SWS profiles for each of the primary bands (left) and the index profiles (right). 
    Third row: The reflectance spectra for each primary band and sub-bands (blue and orange shades) calculated with TMM (red) and RCWA (Green). 
    }  
\label{fig:design_optim_noloss}
\end{figure}

We calculate the effective index, $n(z)$ for a range of pairs $(d, \Gamma_{\text{m}})$. By convention, $n(0) = n_{\text{air}} = 1.0$ and $z$ increases toward the substrate such that $n(d) = n_{\text{sub}}$. For each $n(z)$ we use the transfer matrix method~\cite{hecht_opt} to calculate the expected reflectance $R(\nu)$, and the band averaged reflectance $\bar{R}_{l}$ and $\bar{R}_{h}$ for the lower and higher frequency sub-bands, respectively. 
We make maps of $\bar{R}_{l}$ and $\bar{R}_{h}$ as a function of 
$(d, \Gamma_{\text{m}})$, see the top row of Figure~\ref{fig:design_optim_noloss}.
For each primary band we choose the structure height as the minimum value of $d$ for which both $\bar{R}_{l}$ and $\bar{R}_{h}$ are smaller than 0.02 and  denote it $d_{\text{opt}}$. 
In most cases the value of $d_{\text{opt}}$ determines a single $\Gamma_{\text{m},\text{opt}}$. 
We have determined that when several $\Gamma_{\text{m}}$ values are obtained for $d_{\text{opt}}$ they yield negligible differences in the actual physical shape. The values $d_{\text{opt}}$ are given in Table~\ref{tab:designed_pitch_depth}.

The optimal values $d_{\text{opt}}$, $\Gamma_{\text{m},\text{opt}}$ determine an optimal profile $n_{\text{opt}}(z)$ -- see Figure~\ref{fig:design_optim_noloss}, second row,  right panel --  and we convert this profile to a physical shape using second-order effective medium theory (EMT)~\cite{1994Brauer}. The profile is split to 200 consecutive layers and at each layer we use EMT to find the appropriate alumina fill fraction that would produce the desired $n$. The set of 200 fill fractions determines a physical shape. The resultant physical shapes are given in Figure~\ref{fig:design_optim_noloss}, second row, left panel. 

We calculate the expected reflectance of the designed physical profiles using rigorous coupled wave analysis (RCWA)~\cite{Moharam:86,Moharam:95} \footnote{RCWA calculations were carried out with DiffractMOD, Synopsys, Inc.}. This reflectance is compared to the one predicted by the Klopfenstein $n(z)$ profile using TMM and the 200 index layers that were used earlier to find the fill fractions. The two reflectance spectra are given in the third row of Figure~\ref{fig:design_optim_noloss}, and the average reflectances are given in Table~\ref{tab:designsummary}.

\begin{table}[ht] 
    \centering
    \caption{The averaged reflectance at each band calculated from the TMM- and RCWA-based spectra. We assume a top-hat band shape.}
    \begin{tabular}{c c c c c c c}
            \hline
          Primary band & \multicolumn{2}{{c }}{LF} & \multicolumn{2}{ c }{MF} & \multicolumn{2}{ c}{HF}   \\ \hline
         Frequency band [GHz]& 27  & 39  & 93 & 145 & 225 & 280  \\ \hline \hline
     EMT    & 0.02 & 0.01 & 0.02 & 0.00 & 0.02 & 0.01  \\
     RCWA   & 0.02 & 0.01 & 0.02 & 0.00 & 0.01 & 0.01  \\
     \hline
    \end{tabular}
    \label{tab:designsummary}
\end{table}

\begin{figure}[ht]

\begin{center}        
\includegraphics[width=0.4\textwidth]{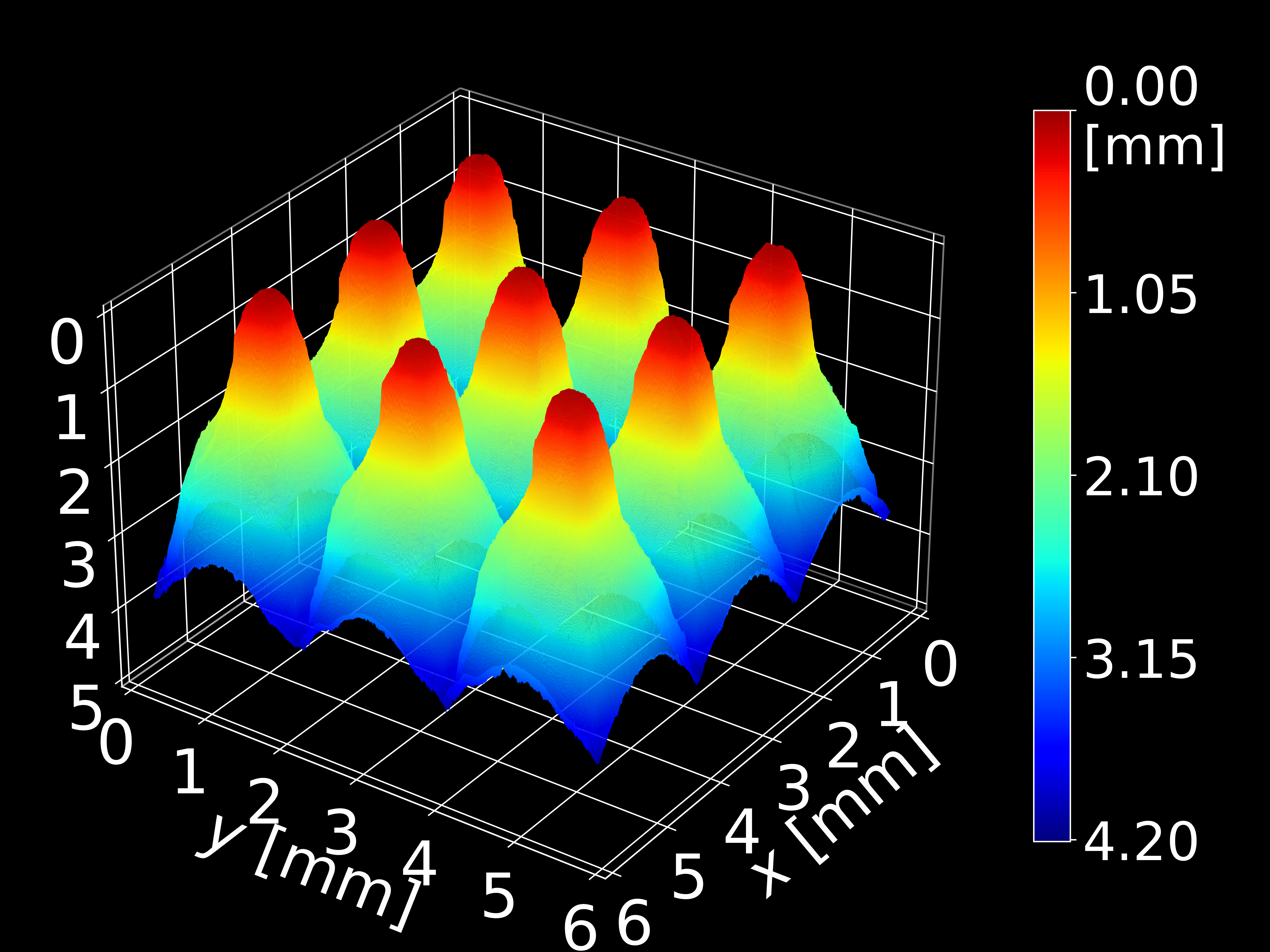}
\includegraphics[width=0.35\textwidth]{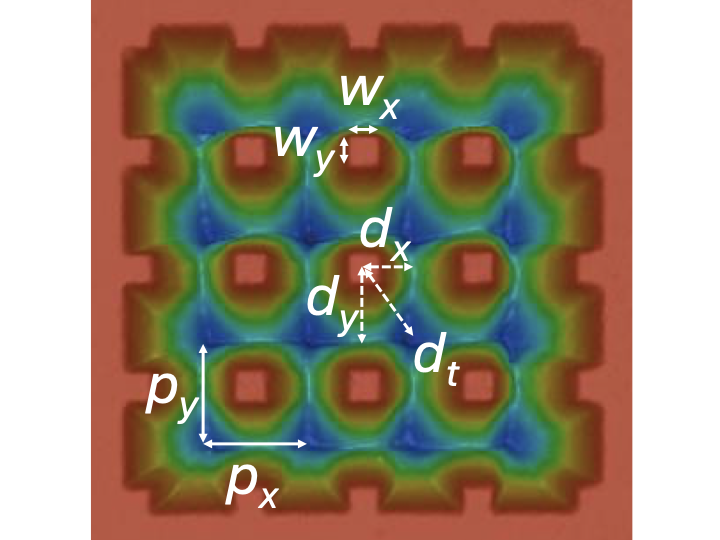} 
\end{center}
\caption{Left: A confocal microscopy image of a 3$\times$3 SWS for LF. 
Right: Top view of SWS with annotation indicating the shape parameters characterizing the SWS. 
Depth measurements are shown as dashed arrows.}
\label{fig:vk4_iso}
\end{figure}



    \begin{table}[ht]
         \caption{The parameters used for the laser machining.}
         \centering
         \begin{tabular}{  c  c  }  \hline
         \multicolumn{2}{c}{Laser model - Light Conversion Carbide CB3} \\ 
         \hline
         Wavelength & 1030 nm \\
         Repetition rate & 100 kHz \\
         Pulse energy & 100 -- 400 \textmu J \\
         Pulse duration & 4 ps \\
         Spot diameter (1/$e^{2}$) &  30.2 \textmu m \\ 
  \hline
         \end{tabular}
         \label{tab:laser_config}
    \end{table}

\section{Fabrication and Shape Measurements}
\label{sec:fabrications}

\begin{figure}[ht]
    
    \centering
    \begin{subfigure}{0.48\textwidth}
        \centering
        \includegraphics[width=\textwidth]{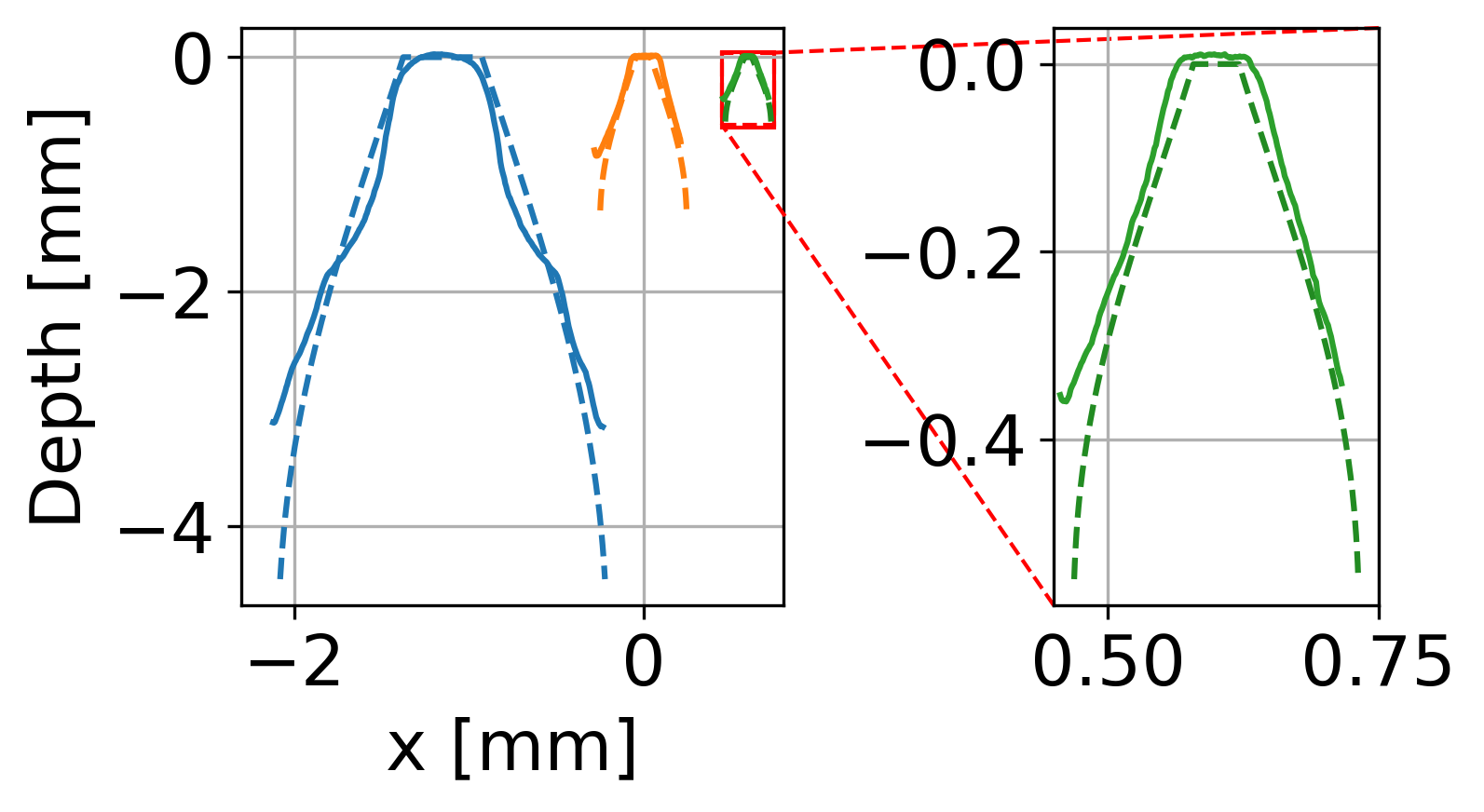}
        \caption{Saddle depth $x$}
        \label{fig:cross_x}
    \end{subfigure}
    \hfill
    \begin{subfigure}{0.48\textwidth}
        \centering
        \includegraphics[width=\textwidth]{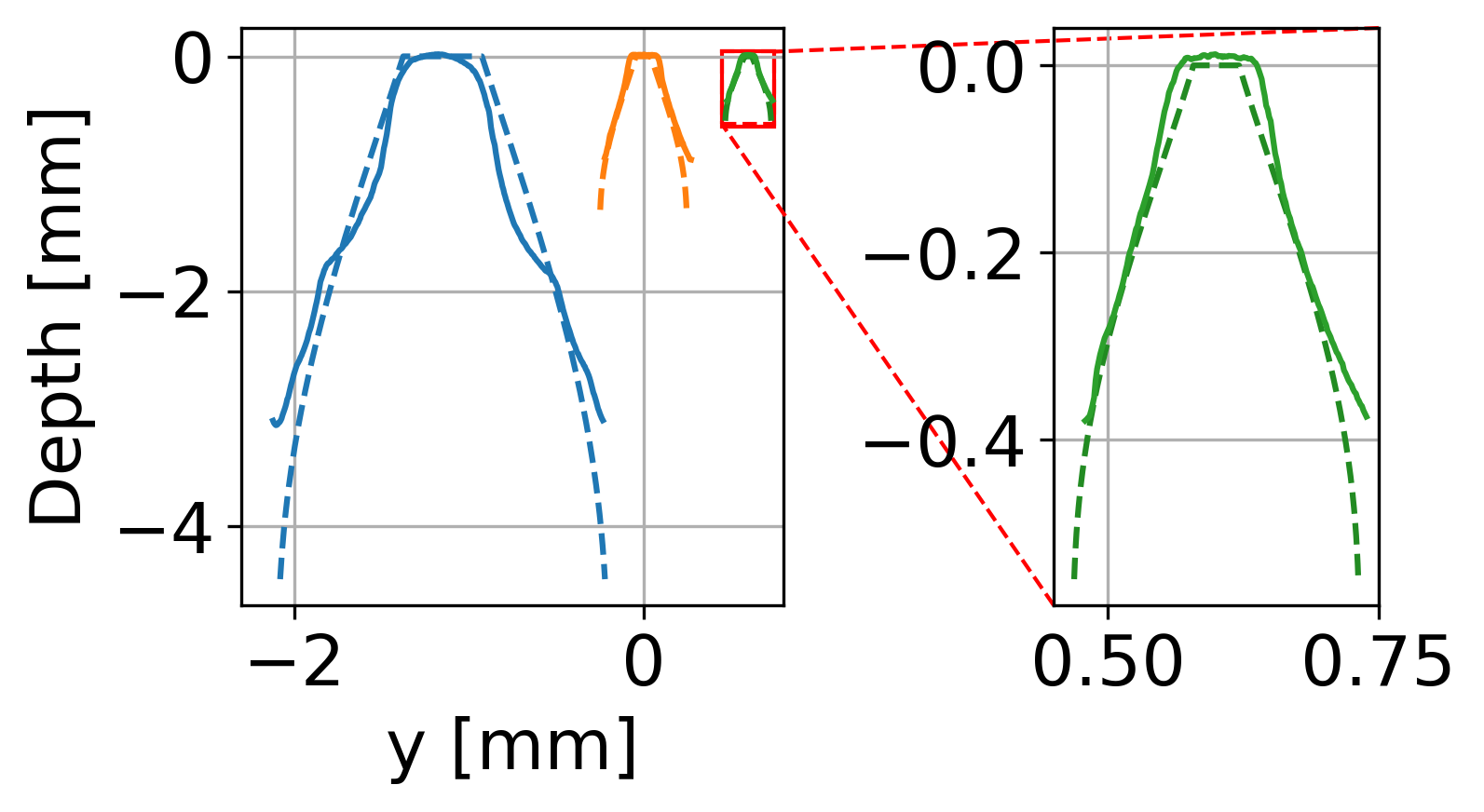}
        \caption{Saddle depth $y$}
        \label{fig:cross_y}
    \end{subfigure}
    \begin{subfigure}{0.48\textwidth}
        \centering
        \includegraphics[width=\textwidth]{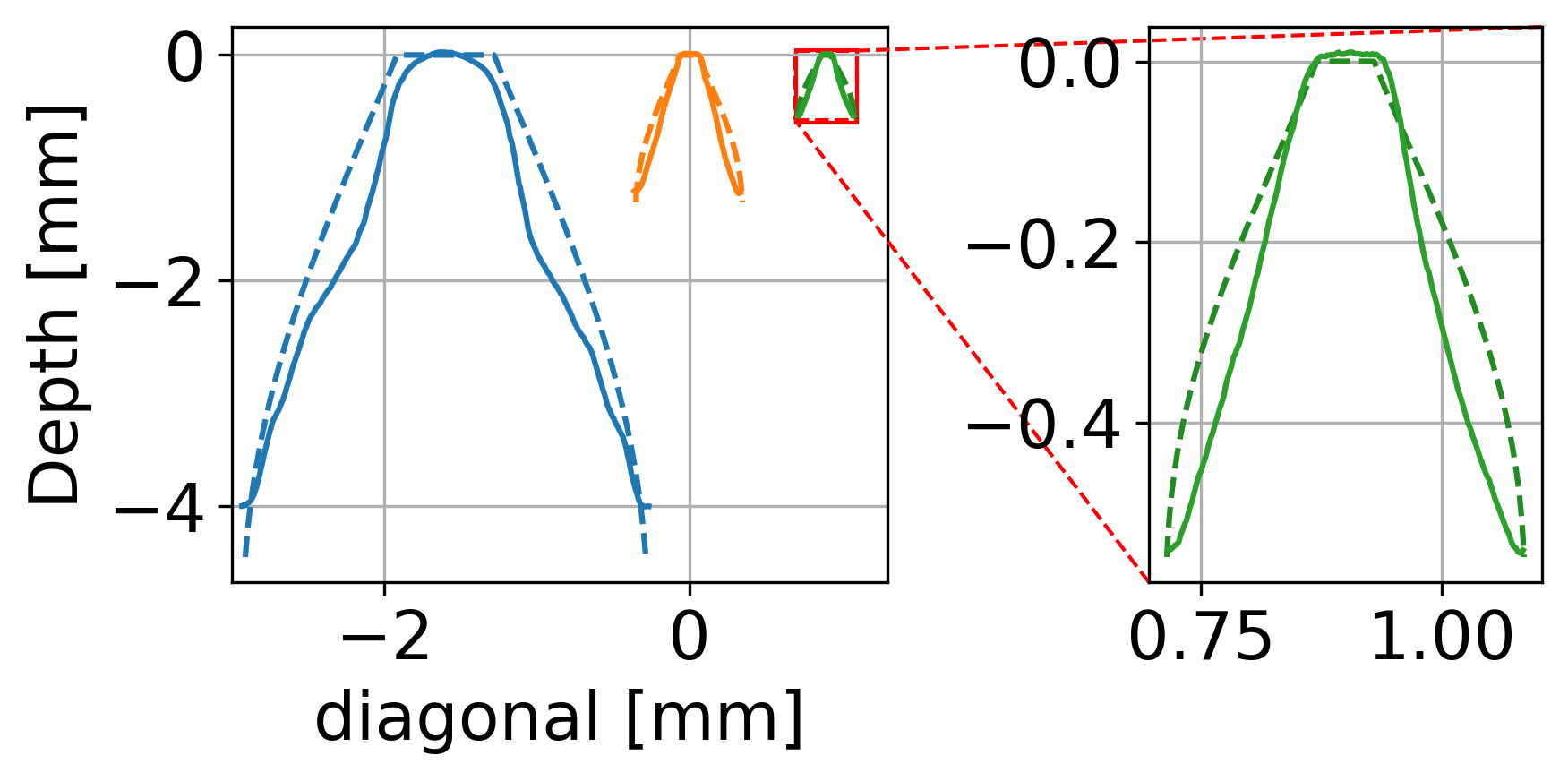}
        \caption{Total depth}
        \label{fig:cross_diag}
    \end{subfigure}
    \hfill
    \begin{subfigure}{0.45\textwidth}
        \centering
        \includegraphics[width=\textwidth]{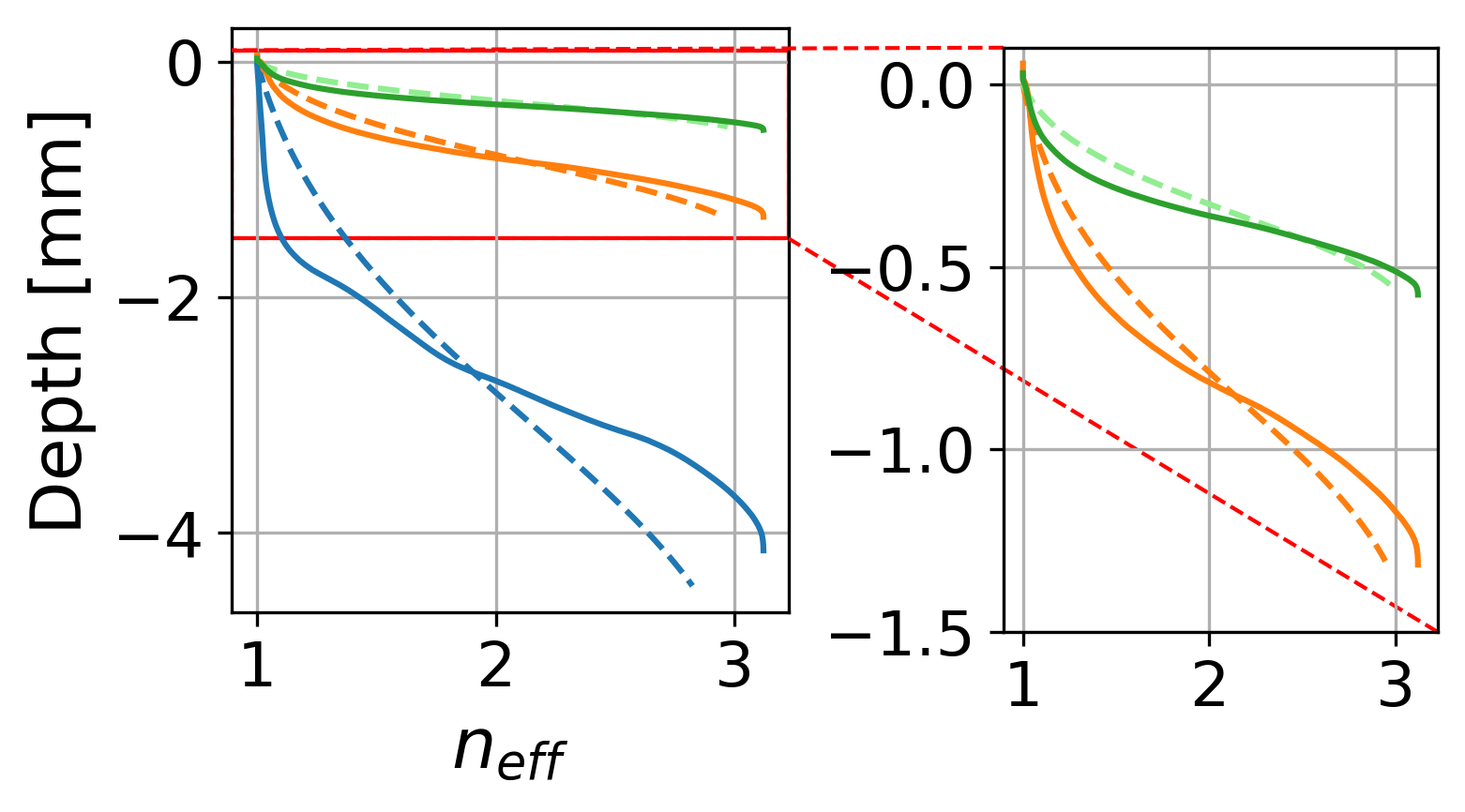}
        \caption{Index profile}
        \label{fig:index_profile}
    \end{subfigure}
    \hfill
    \begin{subfigure}{0.35\textwidth}
        \centering
        \includegraphics[width=\textwidth]{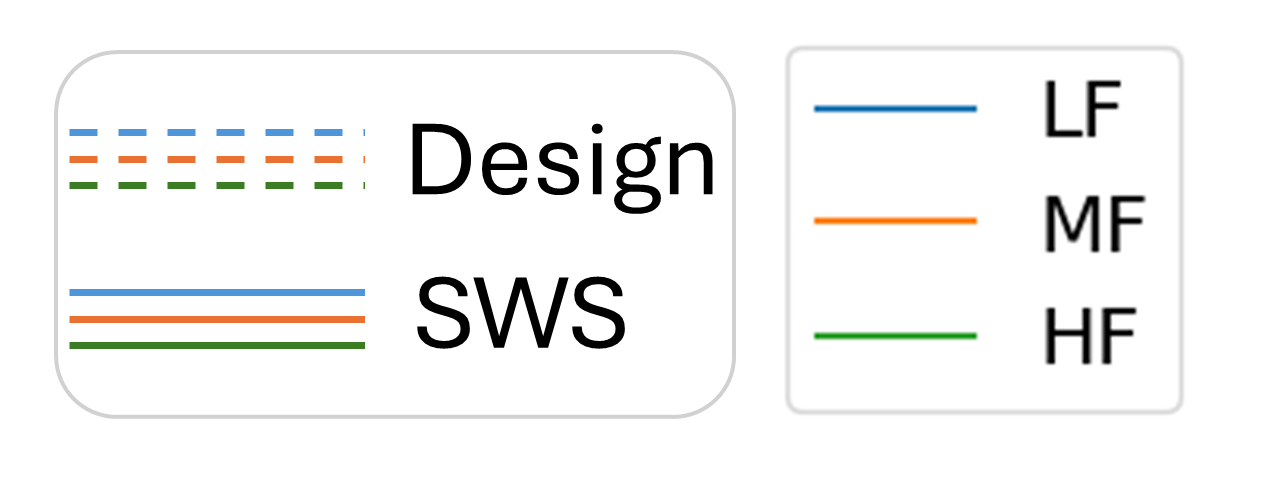}
        \label{fig:crosssection_legend}
    \end{subfigure}
    \hfill

    \caption{Panels (a)-(c): cross-sections of the central fabricated pyramids and designed shapes (solid and dash) for each of $x$, $y$, and diagonal directions. Each panel gives the cross sections for all bands. 
    Panel (d): Index profiles of fabricated shapes and the designed shapes (solid and dash). 
    }
    \label{fig:crosssection}
\end{figure}

For each of the three designs, we use laser ablation to fabricate a $3\times3$ SWS array on alumina.\footnote{The material is 99.5LD from CERATEC CO., LTD.} The laser parameters are listed in Table~\ref{tab:laser_config}. The laser beam is scanned across the surface to produce two dimensional grooves in a manner similar to earlier samples
produced by our collaboration~\cite{matsumura2016,Schutz2016,Young2017,matsumura2018,Sakurai2019,takakuJAP2020,takakuSPIE2020,wen2021,mustang2,Takaku2022,Golec:22,Takaku2023}. 
The left panel in Figure~\ref{fig:vk4_iso} shows confocal microscopy images of the SWS made for LF.\footnote{Confocal microscopy was done with VK-X1000, KEYENCE Inc.} Similar quality SWS have been obtained for other bands. We use the image files to evaluate structure shape parameters including depths between pyramids along the $x$ and $y$ dimensions ($d_x$ and $d_{y}$), diagonal depth ($d_t$), pitches ($p_x$ and $p_y$) and tip dimensions ($w_x$ and $w_y$). These parameters are shown in the right panel of Figure~\ref{fig:vk4_iso}. Table~\ref{tab:shape_parameters} gives the measurements of the shape parameters. 
Figure~\ref{fig:crosssection} shows cross-sectional views of the center fabricated pyramids and the designed shapes, and a comparison of the designed index profiles to the ones derived from the measured shapes. To find a fabrication-based index profile we split the $z$ data of each measured $3 \times 3$ pyramid array to 200 equal thickness layers. For each layer $i$
we find a linear fill fraction $f_{i}$ by taking the root of the area fill fraction, and we use EMT and $f_{i}$ to find an effective index of refraction $n_{i}$. The profile in Figure~\ref{fig:crosssection} is the combination of the 200 $n_{i}$.

\begin{table}[ht]
 \centering
 \caption{Averages and standard deviations of SWS shape parameters. The averages were obtained from nine pyramids. 
 }
 \begin{tabular}{  c  c  c  c  }  \hline
   (units: mm)    & LF  & MF  & HF  \\ 
 \hline \hline
 pitch $x$ ($p_x$) & 1.89 $\pm$ 0.01 & 0.50 $\pm$ 0.00 & 0.26 $\pm$ 0.00  \\
 pitch $y$ ($p_y$) & 1.90 $\pm$ 0.02 & 0.50 $\pm$ 0.01 & 0.26 $\pm$ 0.01  \\
 \hline
 Saddle depth $x$ ($d_{x}$) & 3.19 $\pm$ 0.03 & 0.88 $\pm$ 0.02 & 0.38 $\pm$ 0.01  \\
 Saddle depth $y$ ($d_{y}$) & 3.18 $\pm$ 0.02 & 0.90 $\pm$ 0.01 & 0.40 $\pm$ 0.01 \\
 Total depth ($d_{t}$) & 4.13 $\pm$ 0.03 & 1.31 $\pm$ 0.02 & 0.58 $\pm$ 0.01 \\
 \hline
 Top width $x$ ($w_x$) & 0.43 $\pm$ 0.03 & 0.14 $\pm$ 0.01 & 0.06 $\pm$ 0.00 \\
 Top width $y$ ($w_y$) & 0.42 $\pm$ 0.03 & 0.15 $\pm$ 0.01 & 0.05 $\pm$ 0.00 \\
 \hline
 \end{tabular}
 \label{tab:shape_parameters}
\end{table}

\section{Predicted Performance } \label{sec:results}

\subsection{Reflectance Spectrum}

\begin{figure}[ht]
    \begin{center}

    \includegraphics[width=1.0\textwidth]{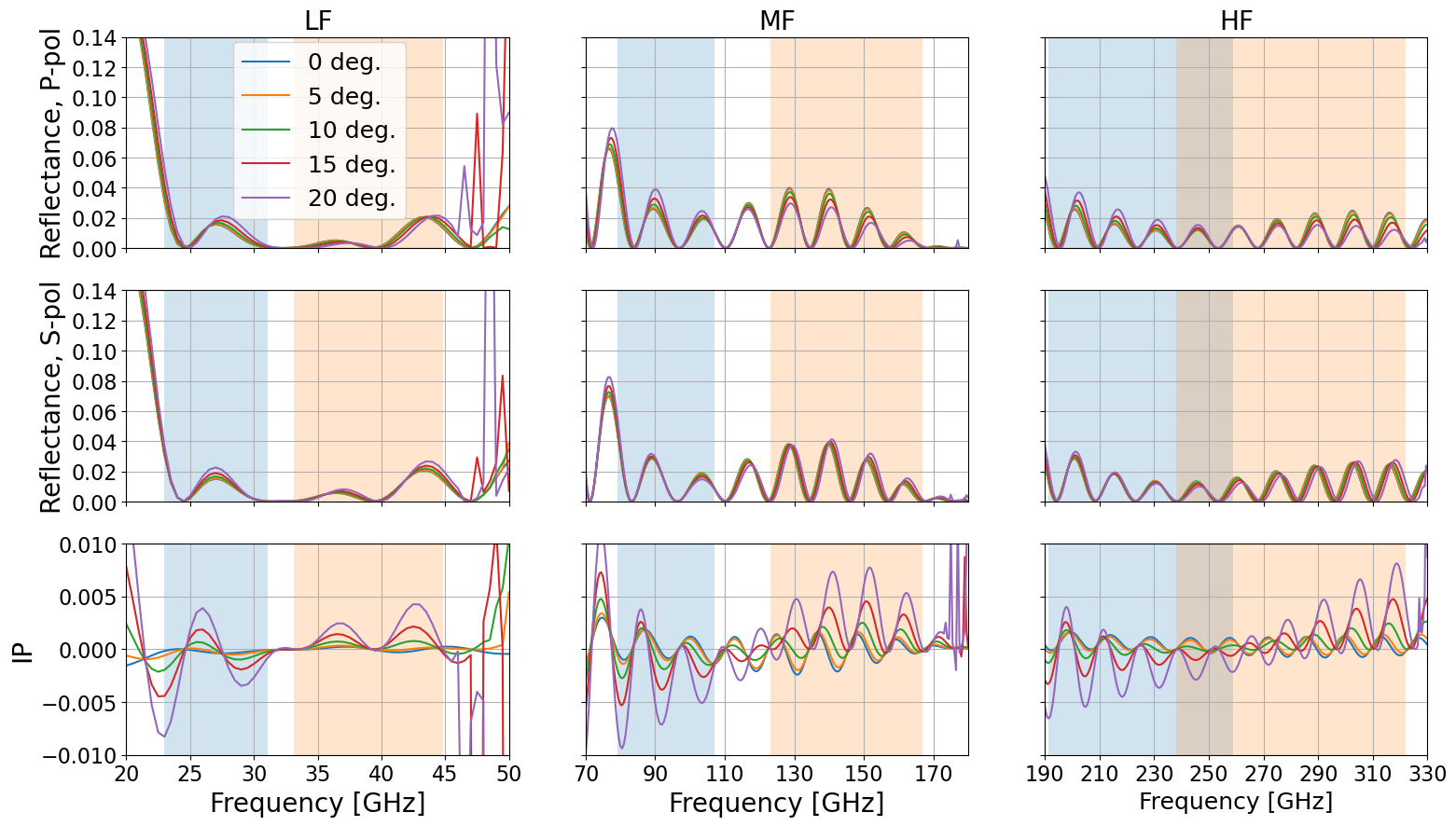}
    \end{center}
    
    \caption{ Predicted reflectance spectra and instrumental polarization for filters with the fabricated SWS-ARC as a function of incidence angle and polarization state of the incident light.}
\label{fig:transmittance_validation}
\end{figure}

To calculate the reflectance spectra of filters with SWS-ARC we construct filter models that have the fabricated shapes on both sides of a 3~mm thick alumina. For example, the total distance from tip to tip of an HF filter would be 4.1~mm, see Table~\ref{tab:shape_parameters}.  
We assume a periodic array of the central pyramid of each fabricated SWS topology, and that the $x$ orientation in one side is perpendicular to $x$ in the other side of the disc such that fabrication-induced $x,\, y$ asymmetries tend to cancel. This is straightforward to achieve in practice.~\cite{mustang2} We produce spectra using RCWA for a range of incidence angles and in two linear polarization states, see Figure~\ref{fig:transmittance_validation}. We average the reflectance spectra and find that for all four sub-bands of LF and HF, for all incidence angles up to 
20 degrees, and in both polarization states, the average reflectance is 1\%. For MF, all values are 2\% or less; more than 60\% of the values are 1\%.



\subsection{Instrumental Polarization}

We compute the instrumental polarization ($IP$) defined as 
\begin{eqnarray}
    \label{eq:IP}
    IP(\nu) = \frac{T_s(\nu)-T_p(\nu)}{T_s(\nu)+T_p(\nu)},
\end{eqnarray}
where $T_p$ and $T_s$ are the transmittance of $p$ and $s$-polarized incident light, respectively. No absorption is included and therefore $T = 1-R$, where $R$ is the reflectance. The average IP per band as a function of incidence angle is given in Table~\ref{tab:predictedperformance_ip}. We find that the largest average IP is 0.3\% and in most cases it is less than 0.1\%.

\begin{table}[ht] 
    \centering
    \caption{Average instrumental polarization as a function of incidence angle. } 

    \begin{tabular}{c | c c c c c c} 
          \hline
          &  \multicolumn{6}{c}{$\bar{IP}$ ($\boldsymbol{\times 10^{-4}}$)}  \\
          \hline
         Incidence &  \multicolumn{2}{c }{LF} & \multicolumn{2}{c }{MF}  & \multicolumn{2}{c}{HF}  \\
         angle (deg)  & 27 GHz  & 39 GHz  & 93 GHz & 145 GHz & 225 GHz & 280 GHz  \\ \hline\hline
        
          0   & 1.9 & 0.1 & 2.0 & 1.5 & 5.7 & 2.7 \\ 
         5     & 2.5 & 1.0 & 0.4 & 0.6 & 4.7 & 3.5 \\
         10    & 4.3 & 3.9 & 4.9 & 6.6 & 1.4 & 5.8 \\
         15    & 7.7 & 9.0 & 14.6 & 16.5 & 4.6 & 9.0 \\
         20    & 13.4 & 16.8 & 29.9 & 29.9 &14.3 & 12.0 \\
         \hline
    \end{tabular}

    \label{tab:predictedperformance_ip}
    \end{table}

\section{Discussion and Summary} \label{sec:discussions}


We designed SWS-ARC for the three primary bands of the SO-SAT. The fractional bandwidths for these ARC exceed 50\% and reach up to 72\%, see Table~\ref{tab:band}. Because the analysis in this paper focuses on minimizing reflectance, we required that the in-band average reflectance would be equal to or less than 2\%. 
The design of an actual filter for the SO-SAT, or for other instruments, should also take into account absorptive losses, which may change the chosen ARC-SWS topologies. When accounting for both absorptive and reflective losses one maximizes transmission rather than minimizing reflections, and the overall thickness of the sample becomes an adjustable parameter. The thickness is determined by both mechanical robustness considerations as well as loss. When absorptive loss dominates the overall loss, there is no requirement to minimize reflections. 

The small-area prototype SWS-ARC give shape parameters that when extrapolated to large area give predicted reflectance values equal to or less than 2\%, consistent with the initial designs. For the specific case of the SO-SAT, there is a relatively good balance between absorption and reflection. Assuming a room temperature loss $\tan\delta < 5.0\times10^{-4}$~\cite{Inoue2016, mustang2}, we find a loss of $\sim$1.5\% with a substract thickness of 3~mm. The absorption would be somewhat higher when accounting for the SWS. 

We calculated reflection spectra as a function of incidence angle and found very weak dependence for angles up to 20~degrees. We found that when the effects of IP are averaged over all incidence angles converging into any focal plane position the net IP is predicted to be smaller than 0.3\%. 

Laser ablation can be considered an effective technique for making SWS-ARC only if the time and cost to make them are reasonable. The time to make a laser ablated SWS-ARC is determined by the volume removal rate (VRR)~\cite{wen2021}, and the VRR for the three prototypes presented in this paper are given in Table~\ref{tab:process_rate}. The VRR for the LF is comparable to the rate we demonstrated in earlier fabrications~\cite{mustang2,wen2021}. More research is required to improve the rates for the MF and HF. 
 
\begin{table}[ht]
    \centering
    \caption{The volume removal rate ($VRR$) at each band.}
    \begin{tabular}{ c  c  c  c } 
    \hline
    & LF & MF & HF \\
    \hline
    $VRR$ [$\mathrm{mm^{3}}$/min.] & 19.6  & 8.4  & 4.6   \\
    \hline
    \end{tabular}
    \label{tab:process_rate}
\end{table}


\section*{Disclosures}
The authors declare there are no financial interests, commercial affiliations, or other potential conflicts of interest that have influenced the objectivity of this research or the writing of this paper.

\section*{Data Availability}
The image data is proprietary.

\section*{Acknowledgments} 

We acknowledge the World Premier International Research Center Initiative(WPI) for support through Kavli IPMU. 
This work was supported by JSPS KAKENHI Grant Number JP23H00107. 
This study was funded in parts by MEXT Quantum Leap Flagship Program (MEXT Q-LEAP, Grant Number JPMXS0118067246). 
Part of this work was supported by NSF grant numbers NSF-2206087 and NSF-2348668.
This work was supported by JST SPRING, Grant Number JPMJSP2108. 
This work was supported by UTokyo Foundation and Iwadare Scholarship Foundation. 
This paper is based on the SPIE proceeding~\cite{Aizawa2024}.


\bibliography{report} 
\bibliographystyle{spiebib} 

\end{document}